      [1999/12/01 v1.4c Il Nuovo Cimento]
\documentclass{cimento}
\newcommand{\ve}[1]{\mbox{\boldmath$ #1 $}}

\usepackage{graphicx}  
\title{Self-consistent radiative effect on relativistic electromagnetic particle acceleration}
\author{K.~Noguchi\from{ins:rice}\ETC,
E.~Liang\from{ins:rice},
K.~Nishimura\from{ins:adv}}
\instlist{\inst{ins:rice} Dept. of Physics and Astronomy, Rice Univ., Houston TX, USA
  \inst{ins:adv} AdvanceSoft, Tokyo Japan}
\PACSes{\PACSit{52.65.-y}{Plasma simulation}
\PACSit{52.65.Rr}{Particle-in-cell method}
\PACSit{52.30.-q}{Plasma dynamics and flow}
}
\begin{document}

\maketitle

\begin{abstract}
We study the radiation damping effect on the relativistic acceleration
of electron-positron plasmas with two-and-half-dimensional particle-in-cell
(PIC) simulation. Particles are accelerated by Poynting flux via the diamagnetic 
relativistic pulse accelerator (DRPA), and decelerated by the self-consistently
solved radiation damping force. With $\Omega_{ce}/\omega_{pe}\geq 10$,
the Lorentz factor of the highest energy particles reaches $\gamma>100$,
and the acceleration still continues.
The emitted radiation is peaked within few degrees from the direction of Poynting flux
and strongly linearly polarized, which may be detectable in $\gamma$-ray burst(GRB) observations.
We also show that the DRPA is insensitive to the initial supporting currents.
\end{abstract}

One of the unsolved problems in astrophysics is the acceleration of nonthermal 
high-energy particles, whose radiation is observed from pulsars, blazars, 
gamma-ray bursts and black holes. Recently, a new mechanism of relativistic 
nonthermal particle acceleration, called the Diamagnetic Relativistic Pulse 
Accelerator(DRPA), was discovered using multi-dimensional Particle-in-Cell(PIC) 
simulations. When a plasma-loaded electromagnetic pulse expands relativistically,
the self-induced drift current creates ponderomotive trap, which drags only the fast 
particles in the trap and leave slow ones behind. At late times the DRPA reproduces many
of the unique signatures of GRBs, including time profiles, spectra and spectral evolution\cite{lian04}.

When charged particles suffer extreme acceleration, radiation loss and damping can become 
important in the plasma energetics and dynamics. 
However, conventional Particle-in-Cell (PIC) simulations of collisionless plasmas have not 
included radiation effects. 
In this article we report PIC simulation results
using a newly developed 2-1/2-D code that includes 
self-consistent radiation damping.  

In PIC simulations, it is impractical to include high-frequency radiation into the electromagnetic 
field calculation 
because the radiation wavelength is much shorter than the spatial resolution of the fields ($\sim$ Debye length
$\lambda_D \equiv c/\omega_{pe}$, where $\omega_{pe}=\sqrt{4\pi\rho e/m_e}$ is the electron plasma frequency).
Accelerated particles can emit up to the critical frequency $\omega_c=3\gamma^2\Omega_{ce}$,
where $\gamma=E/m_e c^2=1/\sqrt{1-v^2/c^2}$ and $\Omega_{ce}=eB/(m_e c)$ is the electron gyro-frequency.
The ratio of the critical radiation wavelength $\lambda_c$ to $\lambda_D$ is given by
$\lambda_c/\lambda_D=(2\pi\omega_{pe})/(3\gamma^3\Omega_{ce})$, which is $\ll 1$
because $\omega_{pe}/\Omega_{ce}<0.1$ in magnetic-dominated cases and $\gamma\gg 1$.

Instead of calculating the high-frequency component directly, we introduce a radiation damping
force in the form of the Dirac-Lorentz equation \cite{land75}.
The relativistic damping force term $\ve{f}_{rad}$ is given by (see \cite{land75} for detailed calculations)
\begin{eqnarray}
\ve{f}_{rad}&=&\frac{2e}{3\Omega_{ce}}k_{rad}
\left\{\gamma\left[\left(\frac{\partial}{\partial t}+\ve{v}\cdot\nabla\right)\ve{E}
+\frac{\ve{v}}{c}\times\left(\frac{\partial}{\partial t}+\ve{v}\cdot\nabla\right)
\ve{B}\right]\right.
+\frac{e}{mc}\biggl[\ve{E}\times\ve{B}\nonumber\\
&&+\left.\frac{1}{c}\ve{B}\!\times\!(\ve{B}\!\times\!\ve{v})
+\frac{1}{c}\ve{E}(\ve{v}\cdot\ve{E})\right]
\left.-\frac{e\gamma^2}{mc^2}\ve{v}
\left[\left(\ve{E}+\frac{1}{c}\ve{v}\times\ve{B}\right)^{\!\!2}\!\!-\frac{1}{c^2}(\ve{E}\cdot\ve{v})^2\right]\right\},\hspace{3mm}
\label{radf2}
\end{eqnarray}
where $\ve{v}$ is the velocity, and $\ve{E}$ and $\ve{B}$ are the self-consistent 
electric and magnetic fields. Here we introduce a non-dimensional 
factor $k_{rad}$ given by
\begin{equation}
k_{rad}=\frac{r_e\Omega_{ce}}{c}=1.64\times10^{-16}\times B\mbox{(gauss)},
\end{equation}
where $r_e=e^2/(mc^2)$ is the classical electron radius. 
The first square bracket term of the radiation damping force (\ref{radf2}) represents the radiation 
damping due to the ponderomotive force acceleration. 
The third square bracket term is Compton scattering by large scale $(\lambda >\lambda_D)$ 
electromagnetic field
which reduces to Thomson scattering in the classical limit~\cite{rybi79}. 
We should note here that the scattering between high frequency radiation and particles is not 
considered since all 
fields are averaged over Debye length.

The power spectrum analysis \cite{lian04} shows that
our simulation model with $\omega_{pe}/\Omega_{ce}=10$ 
corresponds to the GRBs with initial magnetic energy $\sim 10^{51}$ergs by assuming a
$4\pi$ shell of thickness $10^{12}$ cm and radius $10^{13}$cm.
If GRBs are originated from a region $<10^8$ cm,
this magnetic energy implies an initial B $>10^{13}$ G, which corresponds to $k_{rad}\sim 10^{-3}$.
 
We should restrict ourselves not to reach the quantum-limit, 
$\hbar\Omega_{ce}\sim m_e c^2$ or $B>4.4\times 10^{13}$ G, which corresponds 
to $k_{rad}=7.2\times10^{-3}$, or the formula (\ref{radf2}) fails. 
We choose $k_{rad}$ from zero to $10^{-3}$ in the simulation to enhance the radiation 
effect and $|\ve{f}_{rad}|\tau_{sim}\simeq|\ve{F}_{ext}|\Omega_{ce}^{-1}$ so we can 
see the difference between radiative and non-radiative ($k_{rad}=0$) case within 
the simulation time-scale $\tau_{sim}=O(10^4\Omega_{ce}^{-1})$.

We use the 2-1/2D explicit PIC simulation scheme using the explicit leap-frogging method 
for time advancing \cite{bird85}. 
Spatial grids for the fields are uniform in both $x$ and $z$ directions, 
$\Delta x=\Delta z=\lambda_D$. The simulation domain in the $x\!-\!z$ plane is
$-L_{x}/2\leq x\leq L_{x}/2$ and
$0\leq z\leq L_{z}$ with a doubly periodic boundary condition in both directions.

Following Liang et.~al.~\cite{lian04}, the initial plasma is uniformly distributed 
at the center of the simulation box, $-6\Delta x<x<6\Delta x$ and $0<z<L_z$. 
The background uniform magnetic 
field $\ve{B}_0=(0,B_0,0)$ is applied only in the same region, so that the magnetic field freely 
expands toward the vacuum regions, $x>6\Delta x$ and $x<-6\Delta x$ with accelerating plasma. 
We choose $L_x$ to be long enough so that plasma and EM wave never hit the boundaries 
in the $x$ direction within the simulation time.
The initial temperature of plasma is assumed to be a spatially uniform relativistic Maxwellian, 
$k_BT_e=k_BT_p=1$MeV, where the subscripts $e$ and $p$ refer to electrons and positrons. 

\begin{figure}
\includegraphics[width=\linewidth]{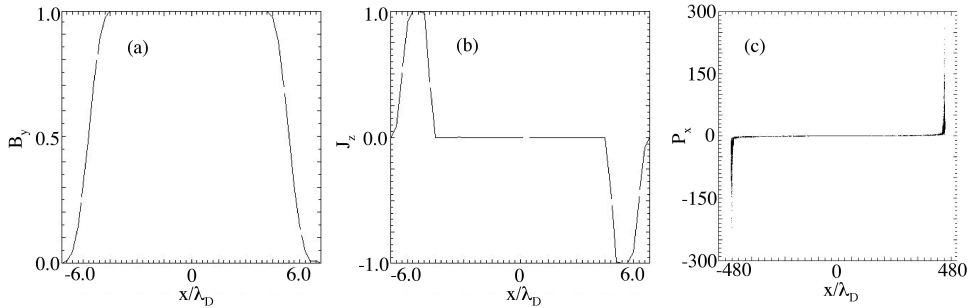}
\caption{\label{fig:fig0} An example of the DRPA with initial current $\ve{J}=\nabla\times\ve{B}$, 
$\omega_{pe}/\Omega_{ce}=0.1$, $k_{rad}=0$. The initial magnetic field and current 
distribution are shown in panels (a) and (b). The phase plot of particles at $t\Omega_{ce}=4800$ is
shown in panel (c).}  
\end{figure}

Since its discovery, questions have been asked if DRPA would persist if the initial static $\ve{B}$
field is confined by supporting currents ($\ve{J}=\nabla\times\ve{B}$), instead of being simply
superposition of opposite traveling waves with initial $\ve{J}=0$. Figure \ref{fig:fig0} shows
an example of a PIC simulation with initial current distribution $\ve{J}=\nabla\times\ve{B}$. 
We use a finer grid $\Delta x=\lambda_D/3$ in this case to resolve the gradient of fields near the plasma surface.
At $t=0$, magnetic field is smoothly decreased toward the edge $x/\lambda_D=\pm6$, and
initial current is locally distributed near the edges. 
However, the initial current is rapidly dissipated by the expanding plasma, and two EM pulses are
formed as in the case without initial current. The asymptotic phase plot is shown
in Fig.~\ref{fig:fig0} (c). It is basically identical to the DRPA phase plot without initial $\ve{J}$.

\begin{table}
\caption{\label{table1}The parameters for radiative runs}
\begin{narrowtabular}{2cm}{cccc}
&$k$&$\omega_{pe}/\Omega_{ce}$& Duration $t\Omega_{ce}$\\
\hline
Run A & 0!         & 0.1! & 10000\\
Run B & $10^{-4}$ & 0.1! & 10000\\
Run C & $10^{-3}$ & 0.1! & 10000\\
Run D & 0!         & 0.01 & 70000\\
Run E & $10^{-4}$ & 0.01 & 70000\\
Run F & $10^{-3}$ & 0.01 & 70000\\
\hline
\end{narrowtabular}
\end{table}

We choose six different sets of parameters shown in Table~\ref{table1} for the radiative 
calculations, by changing 
$k_{rad}=0$, $10^{-4}$, $10^{-3}$ and $\omega_{pe}/\Omega_{ce}=0.1$, $0.01$, 
and run simulations for each case.

\begin{figure}
\begin{center}
\includegraphics[width=0.8\linewidth]{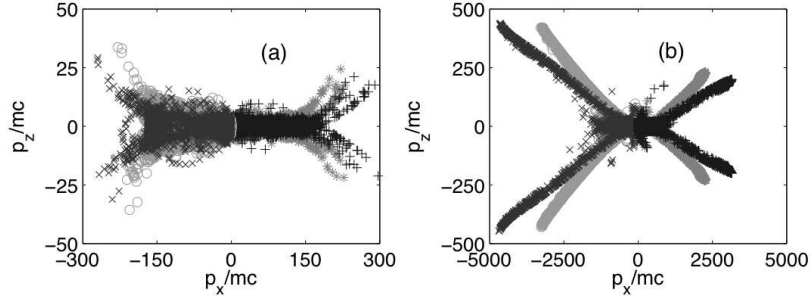}
\end{center}
\caption{\label{fig:phs}Momentum distribution of particles for Run A [(a), $x<0$] and Run C [(a), $x>0$]
at $t\Omega_{ce}=5000$ (circle, asterisk) and $t\Omega_{ce}=10000$ (cross, plus), and
Run .D [(b), $x<0$] and Run F [(b), $x>0$] at $t\Omega_{ce}=35000$ (circle, asterisk) and $t\Omega_{ce}=70000$
(cross, plus). Results in the positive and negative $x$ directions are identical in all cases.}  
\end{figure}

Figure \ref{fig:phs} shows the momentum distribution of particles for (a) Run A($x<0$) and Run C($x>0$), 
and (b) Run D($x<0$) and Run F($x>0$). 
The DRPA accelerates electrons and positrons in the same direction along the $x$ axis, 
whereas electric field accelerates electrons and positrons oppositely
along the $z$ axis, forming X shape distribution in the $p_x-p_z$ plane as a result.
Resulted induced current $J_z$ accelerates particles in the $x$ 
direction by the ponderomotive force $\ve{J}\times\ve{B}$. The ponderomotive force creates
successive 'potential wells' in the $x$ direction, which captures and accelerates co-moving 
particles. We emphasize here that there is no charge separation in the $x$ direction
because there is no mass difference between electron and positron.

Obviously, particle momenta in both $x$ and $z$ directions are radiated away
in both weak and strong magnetic field RD cases. 
For high energy ($\gamma\gg1$) particles, the Compton scattering 
[the third term in Eq. (\ref{radf2})]
becomes dominant damping force, and makes the DRPA acceleration less efficient.

With radiation damping, we can calculate the self-consistent radiation field and 
its angular dependence directly from the velocity and acceleration of each particle.
Intensity $I$ and polarization $\Pi$ of the radiation received by the observer located at $\ve{x}$ 
are given by \cite{jack75,rybi79}
\begin{equation}
I(\hat{\ve{n}},\tau)
=\sum_i\left[
    \left|\ve{E}_i\right|^2
\right]_{\mbox{ret}},\quad \Pi(\hat{\ve{n}},\tau)=\frac{\sqrt{(E_z^2)^2+(E_y^2)^2-2E_y^2E_z^2+U^2}}
    {E_z^2+E_y^2},\label{inten}
\end{equation}
where
\begin{eqnarray}
E_y^2(\hat{\ve{n}},\tau)=\sum_i\left[\left|
    \ve{E}_i\cdot\hat{\ve{y}}
\right|^2\right]_{\mbox{ret}},\quad
E_z^2(\hat{\ve{n}},\tau)=\sum_i\left[\left|
    \ve{E}_i\cdot\hat{\ve{z}}
\right|^2\right]_{\mbox{ret}},\nonumber\\
U(\hat{\ve{n}},\tau)=2\sum_i\left[
    (\ve{E}_i\cdot\hat{\ve{y}})(\ve{E}_i\cdot\hat{\ve{z}})\right]_{\mbox{ret}},\quad
\ve{E}_i=\frac{e}{c}\frac{\hat{\ve{n}}\times[(\hat{\ve{n}}
    -\ve{\beta}_i)\times\dot{\ve{\beta}}_i]}
    {(1-\hat{\ve{n}}\cdot\ve{\beta})^3 R},\label{EyEz}
\end{eqnarray}
where the sum is over all particles along each light cone.
$\hat{\ve{n}}$ is a unit vector in the direction of $\ve{x}-\ve{r}(\tau)$, 
$\ve{\beta}=\ve{v}(\tau)/c$, and $\dot{\ve{\beta}}=d\ve{\beta}/dt$.
We assume that $|\ve{x}|\gg|\ve{r}|$ so that $\hat{\ve{n}}$ is parallel to $\ve{x}$.
The square brackets with a subscript "ret" mean that the quantity in the brackets is 
evaluated at the retarded time $\tau=t-R/c$, where $R=|\ve{x}-\ve{r}|$. 
To specify the direction of the observer with respect to the $x$ axis,
we introduce $\theta$ and $\phi$ as
$\hat{\ve{n}}=(\cos \theta \cos \phi, \cos \theta \sin \phi, \sin \theta).$

\begin{figure}
\begin{tabular}{cc}
\includegraphics[width=0.51\linewidth]{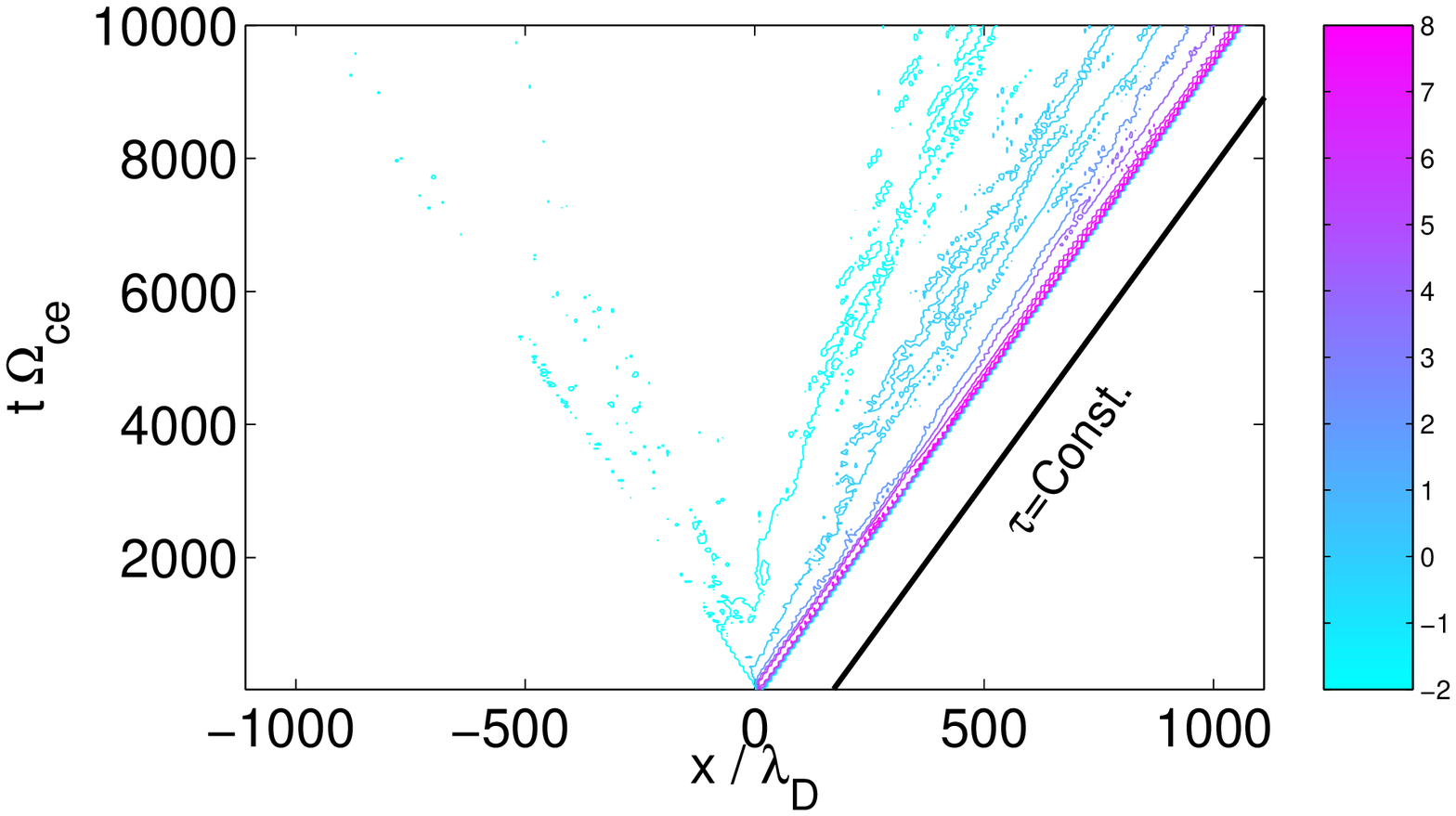}&
\includegraphics[width=0.46\linewidth]{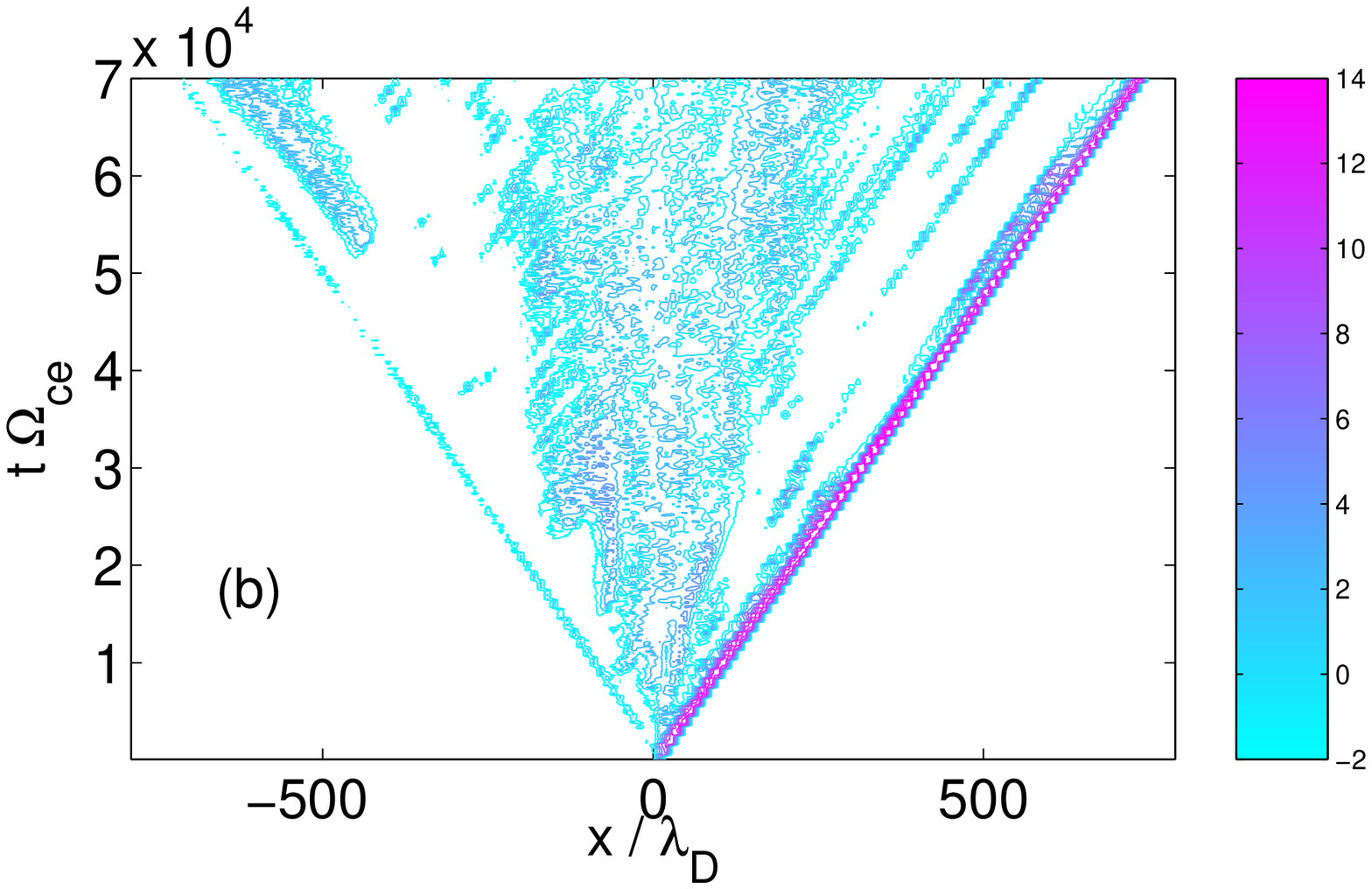}\\
\includegraphics[width=0.49\linewidth]{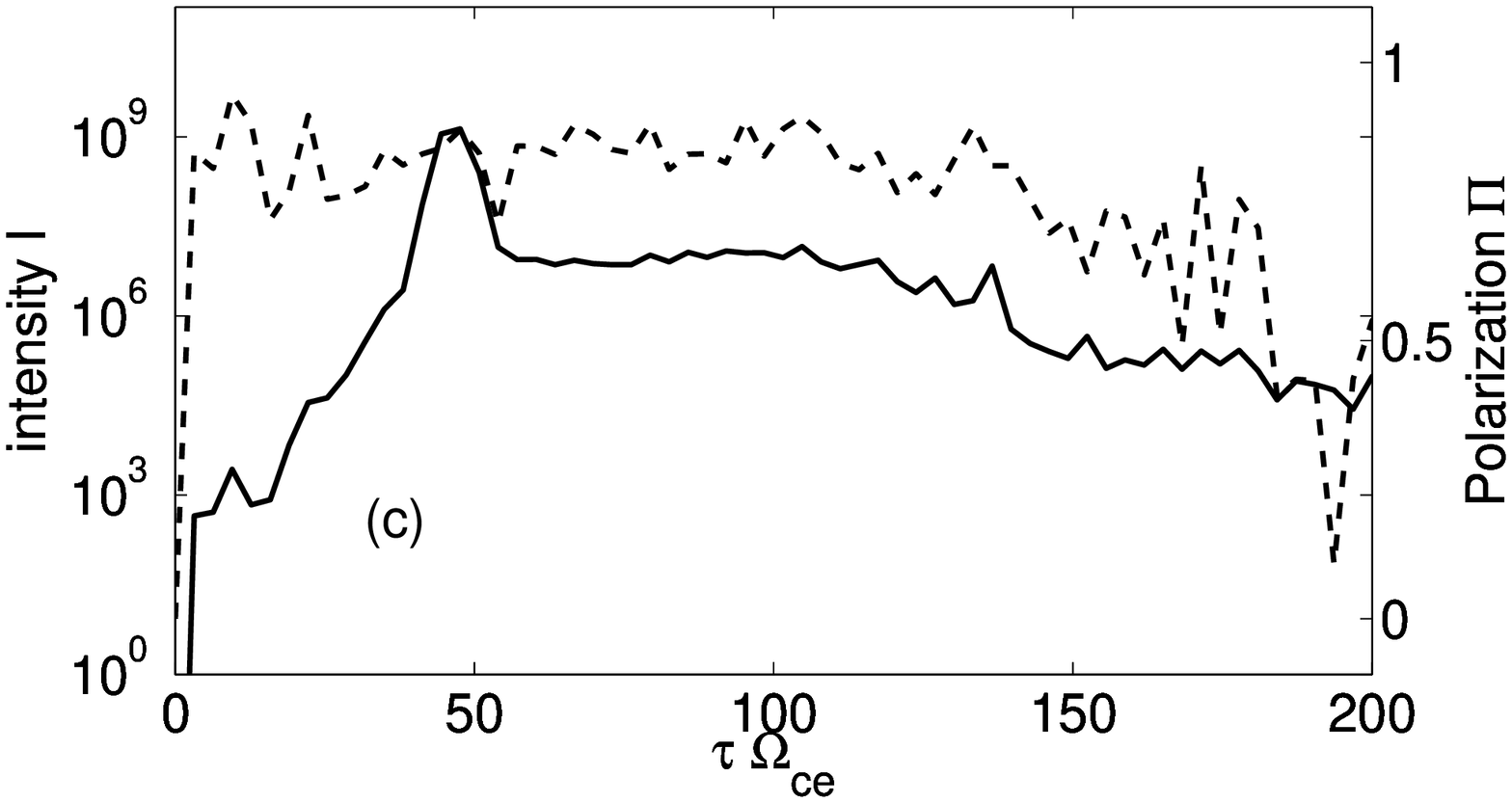}&
\includegraphics[width=0.49\linewidth]{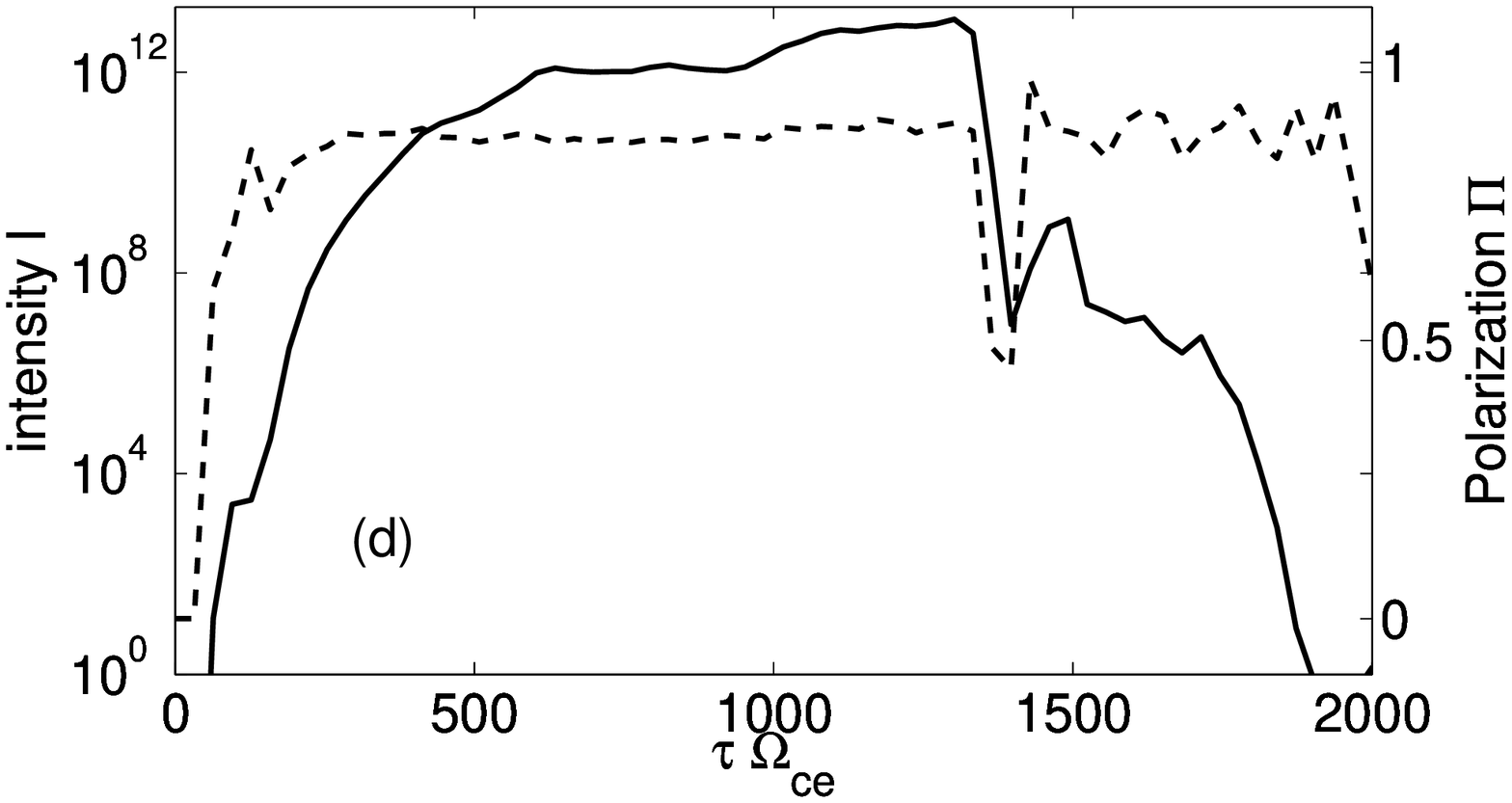}
\end{tabular}
\caption{\label{fig:int} Contour plots of instantaneous intensity $\log_{10} I$ per particle as 
a function of local time $t$ for $\theta=\phi=0$ for (a) Run C and (b) Run F. 
Total detected intensity (solid lines, right scales) and polarization 
(dashed lines, left scales) as functions of observational time $\tau$ for (c) Run C and (d) Run F. 
Intensity is in arbitrary scale. 
To obtain time dependence of intensity, instantaneous intensity from each particle
is summed up along the light cone $\tau=t-R/c=$const., shown as a black line in panel (a). The
light cone moves horizontally leftward with $\tau$.
}
\end{figure}

In Figs.~\ref{fig:int}(a) and (b), 
the contour plot of instantaneous intensity before taking the summation 
over the retarded time is plotted, as a function of local time $t$ with 
$\phi=\theta=0$, illustrating
the ray-tracing technique used in Eqs. (\ref{inten}) and (\ref{EyEz}).
We take a sum of intensity along the light cone $\tau=t-R/c=$const., which is indicated as a solid line
in the panel (a), up to $t\Omega_{ce}=10000$ for Run C and $t\Omega_{ce}=70000$ for Run F.
The light cone moves up in $t$ with increasing $\tau$, and we take $\tau=0$ when the 
pulse front reaches to the observer.
We also note that the intensity is extremely asymmetric because
only energetic particles accelerated in positive $x$ direction can radiate strong emission to the observer.

The time dependence of detected intensity and the polarization with $\phi=\theta=0$ are shown in 
Figs.~\ref{fig:int}(c) and (d). 
Energetic particles
are bouncing back and forth within the ponderomotive potential well, and slower particles are dropped
off to the next well, which broaden the spatial distribution and the resulting radiation duration.

Intensity $I$ is shown as solid lines in Fig.~\ref{fig:int}(b) and (d), indicating the duration time
of strong intensity radiation is $20 \tau\Omega_{ce}$ for Run C,  and $200\sim400\tau\Omega_{ce}$ for Run D with single peak.
Since the driving electromagnetic field is linearly polarized, we expect that the radiation is strongly 
polarized with the small depolarization coming from the initial random $y$ velocity.
Dotted lines in Fig.~\ref{fig:int}(b) and (d) shows that the radiation is strongly linear-polarized as anticipated.

\begin{figure}
\begin{center}
\includegraphics[width=0.5\linewidth]{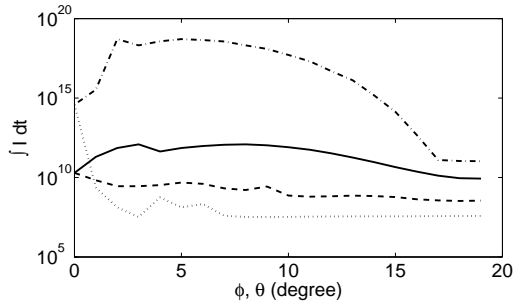}
\end{center}
\caption{\label{fig:itg} 
The total radiation intensity $\int I(\phi, \theta) d\tau$ for Run C and Run F as a function of the 
angle ($\phi, \theta$). Solid: Run C, $\phi=0$; Dashed: Run C, $\theta=0$; Dash-dot: Run F, $\phi=0$;
Dotted: RunF, $\theta=0$.}
\end{figure}

In Fig.~\ref{fig:itg}, we show the total radiation intensity $\int I(\phi, \theta) d\tau$. 
The radiation is a very short 
pulse in Fig.~\ref{fig:int} because we consider single simulation box centered at the origin only.
However, because of the periodic boundary condition in the $z$ direction, 
we should consider the multiple simulation boxes along the $z$ axis, and consider time delay
and angle difference toward the observer from each simulation box, which is very complex
even in the Cartesian coordinate.

Instead of including these geometrical effects into the intensity calculation,
we simply compare the total fluence integrated over $t$ as a function of the angle. 
Angular fluence peaks around $\theta=3\sim 8^\circ$ in $\phi=0$ cases (solid and dash-dot lines), corresponding to 
the direction of high energy particles in Fig.~\ref{fig:phs}. 
Angular fluence rapidly decreases with both $\phi$
and $\theta$, indicating radiation is strongly collimated in the $x$ direction.

Angular fluence distribution in the $\theta$ direction is due to the initial electric field acceleration
in $z$-direction. Since there is no $z$ acceleration after the EM pulse
is formed, eventually all the $z$ momentum will be emitted away, which
narrows the intensity distribution toward positive $x$ direction. At
the same time, however, decoupling of particles from EM pulse in the $x$ direction widens the distribution. 
Thus, the fluence in the $\theta$ direction is always distributed over a finite angle.

In summary, we observed the self-consistent radiation damping effect on the interaction of the EM pulse
with electron-positron plasma via a relativistic PIC simulation. 
Comparison with the non-radiative case showed that radiation
damping force decelerates the energetic particle accelerated by the DRPA, and resulting 
radiation field is strongly linearly polarized within a finite angle from the direction of 
expansion both in weak and strong magnetic field cases,
which may be detectable by $\gamma$-ray burst observations or
laser experiments as an indication of the DRPA mechanism.

The simulations shown here are still too short to determine the radiation 
pattern after particles are completely decoupled from EM pulse. 
Such questions remain as future problems.

\acknowledgments
This research is partially supported by NASA Grant No. NAG5-9223 
and LLNL contract nos. B528326 and B541027. Authors are also grateful to ILSA, LANL, B. Remington and S. Wilks 
for useful discussions.

\end{document}